\documentclass[aps,prx,twocolumn,superscriptaddress,10pt,floatfix,amsmath,amssymb]{revtex4-2}
\usepackage[colorlinks=true,linkcolor=blue,citecolor=blue,urlcolor=blue]{hyperref} 

\usepackage{graphicx} 
\usepackage{dcolumn}
\usepackage{xcolor}
\usepackage{bm}
\usepackage{physics}
\usepackage{lipsum}  
\usepackage{hyperref}
\usepackage{ulem} 

\newcommand{\PauliOp}{\hat{\sigma}}
\newcommand{\HOp}{\hat{H}}
\newcommand{\PAv}{P_{\mathrm{av}}}
\newcommand{\PAvAv}{\overline{P_{\mathrm{av}}}}
\newcommand{\PAvAvX}{\overline{P_{\mathrm{av}}}_X}
\newcommand{\PAvAvZ}{\overline{P_{\mathrm{av}}}_Z}
\newcommand{\GammaOp}{\hat{\gamma}}
\newcommand{\im}{\mathrm{i}}
\newcommand{\Op}{\hat{\mathcal{O}}^{(k)}}

\definecolor{darkblue}{rgb}{0.0, 0.0, 0.55}

\begin{document}

\title{The SYK charging advantage as a random walk on graphs}

\date{\today}
    
\author{Francisco Divi}
    \email{fdivi@perimeterinstitute.ca}
    \affiliation{ICTP South American Institute for Fundamental Research \\
    Instituto de F\'{i}sica Te\'{o}rica, UNESP - Univ. Estadual Paulista \\
    Rua Dr. Bento Teobaldo Ferraz 271, 01140-070, S\~{a}o Paulo, SP, Brazil} 
    \affiliation{Perimeter Institute for Theoretical Physics, Waterloo, Ontario N2L 2Y5, Canada}

\author{Jeff Murugan}
    \email{jeff.murugan@uct.ac.za}
    \affiliation{The Laboratory for Quantum Gravity \& Strings, Department of Mathematics and Applied Mathematics, University of Cape Town, Cape Town, South Africa}
    \affiliation{The National Institute for Theoretical and Computational Sciences, Private Bag X1, Matieland, South Africa}
    
\author{Dario Rosa}
    \email{dario\_rosa@ictp-saifr.org}
    \affiliation{ICTP South American Institute for Fundamental Research \\
    Instituto de F\'{i}sica Te\'{o}rica, UNESP - Univ. Estadual Paulista \\
    Rua Dr. Bento Teobaldo Ferraz 271, 01140-070, S\~{a}o Paulo, SP, Brazil}
    

\begin{abstract}
\noindent
 We investigate the charging dynamics of Sachdev-Ye-Kitaev (SYK) models as quantum batteries, highlighting their capacity to achieve quantum charging advantages. By analytically deriving the scaling of the charging power in SYK batteries, we identify the two key mechanisms underlying this advantage: the use of operators scaling extensively with system size $N$ and the facilitation of operator delocalization by specific graph structures. A novel graph-theoretic framework is introduced in which the charging process is recast as a random walk on a graph, enabling a quantitative analysis of operator spreading. Our results establish rigorous conditions for the quantum advantage in SYK batteries and extend these insights to graph-based SYK models, revealing broader implications for energy storage and quantum dynamics. This work opens avenues for leveraging quantum chaos and complex network structures in optimizing energy transfer processes.
\end{abstract}

\maketitle
\section{Introduction}
The concept of quantum batteries \cite{alicki2013entanglement, campaioli2018quantum, bhattacharjee2021quantum, campaioli2024colloquium}, small quantum systems designed to store and transfer energy, has sparked significant interest as a pathway to overcome classical limitations in energy storage technologies. These systems leverage inherently quantum effects, such as entanglement and coherence, to enhance energy-transfer processes. Over the years, many theoretical aspects have been addressed, including optimal energy extraction protocols \cite{alicki2013entanglement, hovhannisyan2013entanglement, hovhannisyan2020charging, santos2021quantum,safranek2023work}, energy storage protocols \cite{andolina2028charger-mediated, zhang2019powerful, barra2019dissipative, caravelli2020random, tacchino2020charging, quach2020using, crescente2020charging, rodriguez2023catalysis, centrone2023charging,ma2024enhancing, cavaliere2024dynamical, ahmadi2024nonreciprocal, beleno2024laser}, energy storage stability \cite{friis2018precisionwork, rossini2019many-body, santos2019stable, rosa2020ultra-stable}, available energy (the so-called ergotropy) \cite{allahverdyan2004maximal,perarnau-llobet2015extractable, andolina2019extractable, francica2020quantum, shi2022entanglement, yang2023battery, morrone2023daemonic}, different battery architectures \cite{seah2021quantum, shaghaghi2022micromasers, salvia2023quantum,shaghaghi2023lossy, downing2023a, rodriguez2023artificial, morrone2023charging}, and bounds on combinations of figures of merit \cite{garcia-pintos2020fluctuations, cusumano2021comment, garcia-pintos2021reply, bakhshinezhad2024trade-offs, sathe2024universally}. 
In parallel, the first experimental realizations are coming out \cite{hu2022optimal, quach2022superabsorbtion}.

A central question in this field is whether and how quantum systems can charge faster or more efficiently than their classical counterparts, \textit{i.e.} if they can get a quantum advantage in their charging power \cite{binder2015quantacell, campaioli2017enhancing, le2018spin-chain, ferraro2018high-power, andolina2019quantum, crescente2020ultrafast, ghosh2020enhancement, rossini2021quantum, zakavati2021bounds, ghosh2021fast, gyhm2022quantum}. Although early studies established bounds on the achievable advantage of global quantum operations \cite{campaioli2017enhancing, gyhm2022quantum}, very few results are known on the mechanisms that underpin this advantage in concrete quantum systems \cite{mondal2022periodically, gyhm2024beneficial, tirone2024many, demoraes2024quantum}.

The Sachdev-Ye-Kitaev (SYK) model \cite{maldacena2016remarks, polchinski2016the, chowdhury2022Sachdev-Ye-Kitaev}, a paradigmatic example of a strongly interacting, disordered, quantum system, has emerged as a promising platform for exploring such questions. Originally introduced in the context of quantum chaos and holography, the SYK model exhibits features, such as fast scrambling and operator growth, that are ideally suited for studying quantum charging dynamics \cite{rossini2021quantum, kim2022operator, francica2024quantum, romero2024scrambling}. Recent numerical studies have demonstrated the ability of SYK-based batteries to achieve significant quantum advantages in charging power \cite{rossini2019many-body, kim2022operator}. 

In this work, we provide a detailed analytical characterization of the quantum charging advantage in SYK quantum batteries. By recasting the charging process as a graph-theoretic problem, we elucidate the role of operator delocalization and large operator size in driving the advantage. By considering both the integrable and interacting cases, we demonstrate that operator growth is irrelevant in achieving the advantage. Furthermore, we generalize our results to SYK models defined on arbitrary graphs, identifying universal conditions for achieving quantum advantages in these systems.

Beyond their implications for energy storage, our findings shed light on broader questions in quantum many-body physics, including operator spreading and thermalization. This framework bridges the domains of quantum chaos, graph theory, and energy science, providing a robust platform for future explorations into quantum technologies.

\section{SYK Quantum batteries}
We consider batteries made up of $N$ qubits and characterized by a \textit{battery Hamiltonian} $\HOp_0$, which determines its energy content. The system is initially prepared in the ground state of $\HOp_0$, denoted by $\ket{\psi_0}$ and having energy $E_0$. Charging the battery means designing a protocol that evolves the state to a final, more energetic state. A common protocol -- dubbed \textit{double-quench} -- consists of turning on a different Hamiltonian $\HOp_I$ during an interval $0 \leq t \leq \tau$ and letting the system evolve to increase its energy content. In this case, the main freedom in the battery design is determined by the relative choice of $\HOp_I$ with respect to $\HOp_0$.

To determine whether a protocol is successful, figures of merit such as the final energy, the degree of entanglement of the final state, or the stability of the energy stored have been considered and investigated, as outlined in the introduction. Consequently, it is possible to analyze how the charging Hamiltonian $\HOp_I$ should be designed to optimize the particular figure of merit under investigation.

Efficient charging is considered one of the most important features and is probed by the \textit{average charging power} \cite{binder2015quantacell} 
\begin{equation}
    \label{eq:P_average_def}
    \PAv(t) \equiv \frac{E(t) - E_0}{t} \, , \qquad E(t) = \bra{\psi(t)} \HOp_0 \ket{\psi(t)} \, .
\end{equation}
Then, the \textit{optimal charging time} \cite{rossini2021quantum}, $\tau$, is defined as the time that maximizes the average charging power \textit{i.e.} $\tau \equiv \max_t P(t)$.

For systems of $N$ qubits, assuming that both $\HOp_0$ and $\HOp_I$ display a \textit{linear scaling} with $N$ of their bandwidths, it has been shown in \cite{gyhm2022quantum} (see also \cite{campaioli2017enhancing}) that the average power is bounded as,
\begin{equation}
    \label{eq:bound_on_power}
    \PAv(\tau) \leq \alpha N^2 \, ,
\end{equation}
with $\alpha$ being an $N$-independent constant. Moreover, it has been proven that in the absence of entangling operations $\PAv(\tau)$ can scale at most linearly with $N$. The notion of \textit{quantum charging advantage} was therefore coined to indicate any charging protocol for which $\PAv(\tau)$ displays a super-extensive scaling with $N$ \cite{binder2015quantacell, campaioli2017enhancing}. \\

Among the few explicit examples of charging protocols that enjoy a quantum advantage, \textit{SYK quantum batteries} have played a prominent role \cite{rossini2021quantum, kim2022operator, francica2024quantum, romero2024scrambling}. This model is described in terms of $N$ Majorana fermions, $\GammaOp_i$, satisfying $\left\{\GammaOp_i, \GammaOp_j \right\} = \delta_{ij}$, with a Hamiltonian
\begin{equation}
    \label{eq:SYK_hamiltonian_def}
    \HOp_I^{(q)} =  \frac{(\im)^{q/2}}{q!} \sum_{i_1,  \dots , i_q } J_{i_1 \dots i_q} \GammaOp_{i_1} \cdots \GammaOp_{i_q} \, .
\end{equation}
Here, the (fully anti-symmetric) coupling constants $J_{i_1 \dots i_q}$ are randomly extracted from a Gaussian distribution with vanishing mean value and variance $\langle J_{i_1 \dots i_q} J_{i_1 \dots i_q} \rangle = \frac{J^2 (q - 1)!}{N^{q - 1}}$, and where $J$ is an $N$- and $q$-independent constant with the dimension of energy. The integer value $q$ is even, and we will set $J = 1$. 

It should be noted that the Hamiltonians $\HOp_I^{(q)}$ are qualitatively different when $q = 2$ versus the case $q \geq 4$. In the former case, although the model is chaotic from the single-particle point of view, it is integrable as a many-body system. Therefore, no scrambling physics is at work, since the dynamics do not create operator growth. In contrast, the $q\geq 4$ model is many-body and strongly interacting. This sharp difference is reflected in the scrambling properties of the system: for $q \geq 4$ the system is highly chaotic, scrambling, and saturates the Maldacena-Shenker-Stanford bound on the quantum Lyapunov exponent \cite{maldacena2016a, maldacena2016remarks}.  However, the model $\HOp_I^{(2)}$ still enjoys \textit{operator delocalization} \cite{kim2022operator}, \textit{i.e.} by time-evolution a \textit{single} Majorana operator (or a product of $k$ Majorana operators) evolves into a large superposition of Majorana operators (or a superposition of products of $k$ Majorana operators), with the size of the operators involved in the superposition remaining constant.

In this work, we will also consider a \textit{sparse} variant of the SYK$_2$ model, which can be thought of as defined on a fixed graph \cite{xu2020a, garcia-garcia2021sparse, caceres2021sparse, caceres2023out-of-time,garcia-garcia2024sparsity-independent,orman2024quantum}. Specifically, given a graph $G$ with $N$ vertices and a set of edges $E$, we will consider the Hamiltonian 
\begin{equation}\label{eq:SYK_hamiltonian_graph_def}
    \HOp_I^{(2)} =  i \sum_{(i, j) \in E} J_{ij} ~ \GammaOp_{i} \GammaOp_{j} \,
\end{equation}
where the normalization of the antisymmetric coupling constants $J_{ij}$ is given by $\langle J_{ij} J_{ij} \rangle = \frac{N}{2 n_E}$, with $n_E$ being the number of edges of the graph. 

Our main question of interest is: {\it what properties of the graph determine the quantum dynamics of the model?} More precisely, we want to determine whether a $\HOp_I$, defined by a given graph $G$, enjoys a quantum charging advantage. As we will show, the answer to this question is encoded in certain simple graph-related properties.

To this end, we will consider the disorder averaged $\PAvAv$, for a disorder-independent initial state. By computing the time evolution of the battery Hamiltonian, $\overline{\HOp_0 (t)}$, the problem reduces to finding the optimal charging time, $\tau$, and determining whether $\PAvAv(\tau)$ exhibits a super-extensive scaling. If the graph structure is also random, we will average over both disorder and graph configurations.\\

\section{Charging with a quadratic quench as a hopping problem}
\begin{figure}[t!]
    \centering
    \includegraphics[width=0.4\textwidth]{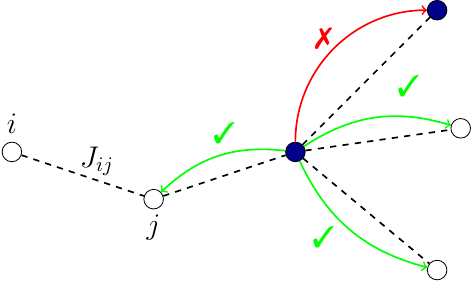}
    \caption{Time evolution is represented as a sum of different paths in the graph of Majoranas. Blue vertices correspond to occupied Majoranas and, as a result of Pauli exclusion, two fermions cannot be at the same vertex in the graph.}
    \label{fig:graph_majoranas}
\end{figure}

As a warm-up, we start by considering the case of the all-to-all connected SYK$_2$, \textit{i.e.} the case in which $G$ is a complete graph. For a single disorder realization, the time evolution of one Majorana is given by $\GammaOp_i (t) = e^{\im \HOp_1 t} \GammaOp_i e^{- \im \HOp_1 t}$ and the Baker–Campbell–Hausdorff (BCH) expansion reads
\begin{equation}
    \label{eq:single_majorana_hopping}
    \GammaOp_i (t) = \sum_{n = 0}^{\infty} \frac{t^n}{n!} \sum_{l_1, \dots , l_n} J_{i l_1} \cdots J_{l_{n - 1}l_n} \GammaOp_{l_n} \, ,
\end{equation} which has a nice geometrical interpretation: each product of couplings $J_{i l_1} \dots J_{l_{n - 1} l_n}$ defines a \textit{random walk} in the graph of Majorana operators, as depicted in Fig. \ref{fig:graph_majoranas}.

Random averaging Eq.~\eqref{eq:single_majorana_hopping} and considering all the Wick contractions greatly reduces the number of paths. First, it implies $i = l_n$, \textit{i.e.} that $\overline{\GammaOp_i}(t)$ is actually computing a \textit{return amplitude}. Second, at the leading order in the large $N$ limit, the average is dominated by \textit{non-crossing} Wick pairings which can be expressed in terms of the \textit{Catalan numbers} and summed to give
\begin{align}
    \label{eq:single_majorana_hopping_averaged_solution}
    \overline{\GammaOp_i} (t) = f(t) \,  \GammaOp_i \, \equiv \frac{J_1(2 t)}{t} \, \GammaOp_i  ,
\end{align}where $J_1(x)$ is a Bessel function of the first kind. Eq.~\eqref{eq:single_majorana_hopping_averaged_solution} reproduces the well-known two-point function of the SYK$_2$ model obtained by summing rainbow diagrams in the large $N$ limit \cite{gross2017a}. 

Eq.~\eqref{eq:single_majorana_hopping_averaged_solution} can be generalized to consider operators of \textit{size} $k$, \textit{i.e.} products of $k$ Majorana operators, $\Op \equiv \GammaOp_1 \cdots \GammaOp_k$ \footnote{Given the complete character of the graph on which the model is defined, the precise choice of the indices $i_1, \dots, i_k$ is immaterial. We therefore assume $i_1, \dots, i_k = 1, \dots, k$.}. 
The time evolution of a size $k$ operator \textit{almost} factorizes to the product of the time evolution of each single Majorana operator
\begin{equation}
    \label{eq:size_k_hopping_averaged_solution}
    \overline{\Op}(t) = f^k \left(t \,  \sqrt{1 - \frac{k}{N}} \right)  \Op \, .
\end{equation} 
The reason why we do not get complete factorization is because two fermions cannot occupy the same site during the random-walk evolution, i.e. we have to take into account a ``Pauli exclusion for Majorana operators'' (see Fig. \ref{fig:graph_majoranas}). For the fully connected graph, this results in a minor correction when $k$ is large but, as we will see in the next section, for certain graphs this can suppress a large number of paths, leading to a \textit{Majorana blockade}.\\

To illustrate these findings in a concrete quantum battery problem, we can compare two seemingly similar battery Hamiltonians
\begin{equation}
    \label{eq:X_Z_model_def}
    \HOp_0^x = \sum_{i = 1}^{N/2} \PauliOp_i^x \, , \qquad \HOp_0^z = \sum_{i = 1}^{N/2} \PauliOp_i^z \, ,
\end{equation}
These models have been dubbed as $X$-model and $Z$-model, respectively \cite{kim2022operator}. Through a Jordan-Wigner transformation, both the Hamiltonians can be expressed in terms of Majoranas (see App.~\ref{app:JW} for the details). The upshot is that, in Majorana language, they show different operator sizes: the $Z$-model is given by a superposition of size-$2$ operators, while the $X$-model is given by a superposition of operators having all possible odd sizes $k = 1, 3, \dots, N - 1$.  

We then can compute the time evolutions of the energy stored in the two batteries using Eq.~\eqref{eq:size_k_hopping_averaged_solution} to find
\begin{equation}
\label{eq:charging_power_X_Z_model}
    \begin{aligned}
    \PAvAvZ (t) &= \frac{N}{2 t} \left(1 - f^2 (t)\right)\,, \\
    \PAvAvX (t) &= \frac{N}{2 t} \left(  1 - \sum_{k=1}^{N/2} f^{2k-1}\left(t\sqrt{1-\frac{2k-1}{N}}\right)\right) \, .
    \end{aligned}
\end{equation}For the $Z$ model, the charging power is maximized for $\tau \approx 1.148 $ giving $\PAvAv_Z(\tau) \approx 0.339 N$,  not achieving a charging advantage.


For the $X$-model, the situation is radically different since it includes operators of size $k \sim N$. To simplify the expression, we replace the sum by an integral in the large $N$ limit and restrict the expression to the early-time dynamics by Taylor expanding:  $f^k\left( t\sqrt{1-k/N}\right) = \exp\left[k \log\left(f \left(t \sqrt{1 - k/N} \right)\right) \right] \approx \exp[-\frac{k}{2} t^2\left(1-\frac{k}{N} \right)] $. The validity of this approximation is given by the extensive scaling of $k \sim N$ that gives a finite result for $t \sim 1/\sqrt{N}$. Consequently, the charging power  
\begin{equation}
\label{eq:charging_power_X-model_SYK_2}
    \PAvAvX(t) \approx \frac{N}{2t}\left( 1 - \int_0^1 \text{d}y ~ e^{-\frac{1}{2}N t^2 y(1-y)}\right) \, ,
\end{equation} which reaches a maximum for $\tau \approx 3.679 ~  N^{-1/2}$, \textit{i.e.} the optimal charging time is \textit{$N$-dependent} and shrinks with increasing the system size.  In turn, the maximum value of the averaged charging power scales as 
\begin{equation}
\label{eq:optimal_charging_power_X-model_SYK_2}
    \PAvAvX(\tau) \approx 0.171 ~ N^{3/2} \, ,
\end{equation}which analytically proves the charging advantage for the $X$-model charged via a SYK$_2$ quench.

\begin{figure}[t!]
\centering
    \includegraphics[width=0.48\textwidth]{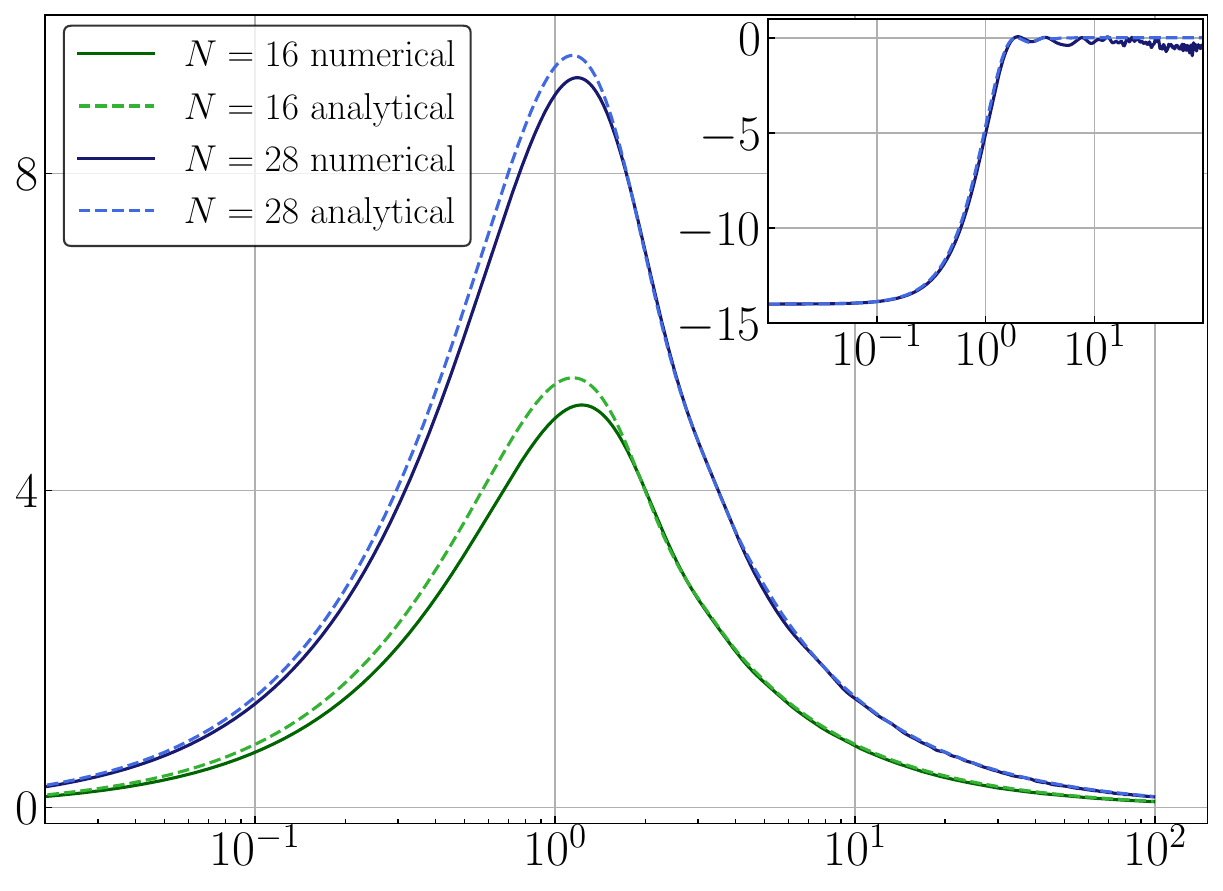}
    \includegraphics[width=0.48\textwidth]{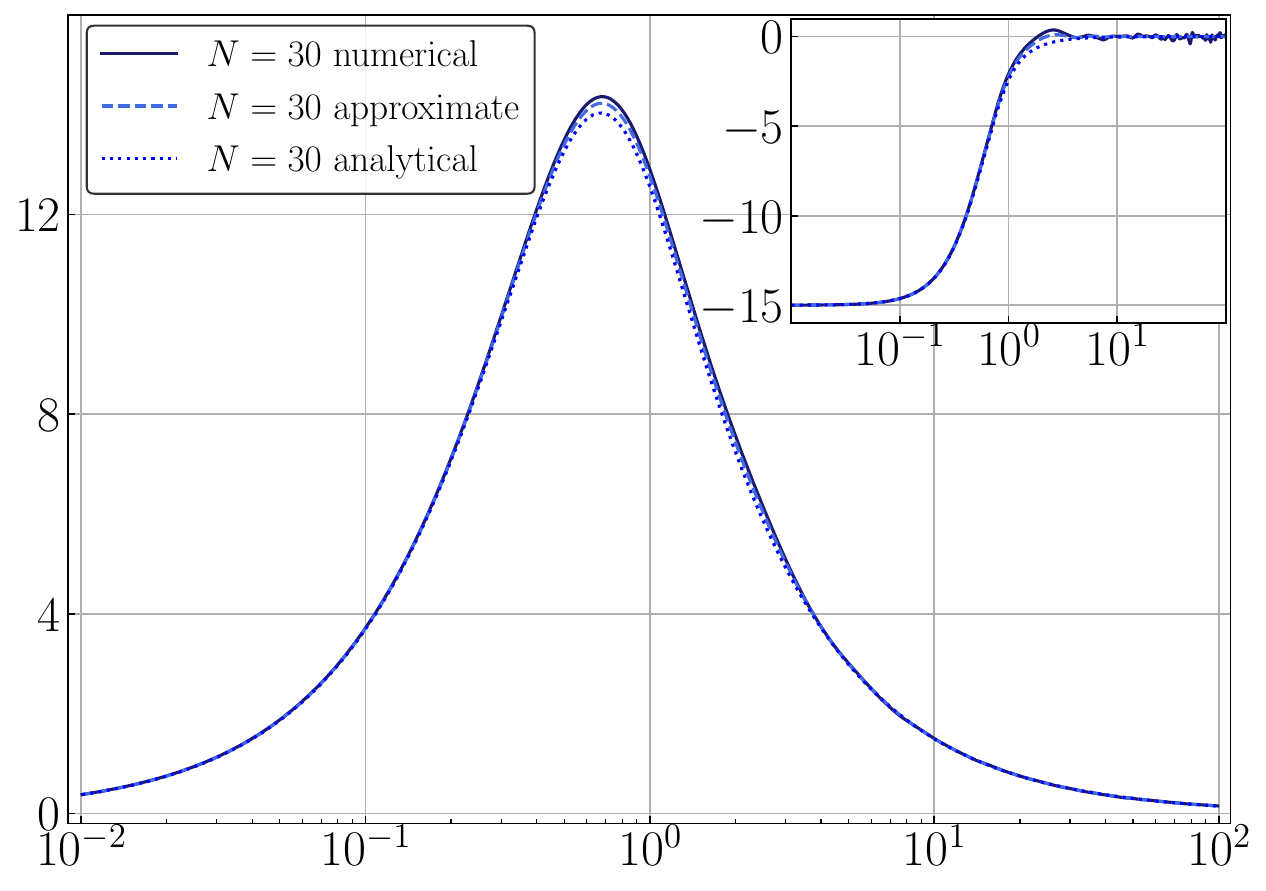}
    \caption{Upper panel: comparison between the analytical formula for the $Z$-model charging power, Eq.~\eqref{eq:charging_power_X_Z_model}, and the numerically obtained charging power for the cases of $N = 16$ and $N = 28$. Lower panel: comparison between the analytical formula for the charging power for the $X$-model, Eq.~\eqref{eq:charging_power_X_Z_model}, its approximate expression, Eq.~\eqref{eq:charging_power_X-model_SYK_2}, and the numerically evaluated power for the case of $N=30$. Insets: the corresponding figures for the average energy, $\overline{E}(t)$, Eq.~\eqref{eq:P_average_def}.}
    \label{fig:charging_power_SYK_2_full}
\end{figure}

The expressions just obtained for the charging power, Eq.~\eqref{eq:charging_power_X_Z_model}, and the approximate expression, Eq.~\eqref{eq:charging_power_X-model_SYK_2}, can be tested against their exact counterparts, obtained via exact time-evolution algorithms on systems of finite dimensionalities. The comparison results are reported in Figs.~\ref{fig:charging_power_SYK_2_full}.
For the $Z$-model the agreement improves by increasing the system size, although deviations are still clearly visible at $N = 28$. On the other hand, for the $X$-model the agreement between the numerical curve and its analytical counterpart is extremely good at $ N = 30$.

This different behavior showed by the two models is related to the different dimensions of the relevant Hilbert spaces explored by the operators during their time evolution (of size $\sim N^2$ for the $Z$-model and $\sim N^{N/2}$ for the $X$-model).

These findings -- the presence/absence of charging advantage and the scaling of the latter -- can be heuristically obtained by path-counting in the graph of Majoranas, Eq.~\eqref{eq:single_majorana_hopping} and its generalization to size $k$ blocks. In the $Z$-model Hamiltonian, each term consists of two Majoranas that can be moved independently; so for each commutator in the BCH expansion the number of paths grows as $\sim N^2$. In general, for \textit{finite} strings of Majoranas, the number of paths grows \textit{polynomially} in $N$ for each commutator. Therefore, $\tau$ has no scaling with $N$ and no charging advantage is obtained.

In contrast, the $X$-model consists of a sum of order $N$-sized operators. Therefore, for each additional commutator, the number of paths grows \textit{super-polynomially} in $N$ like $(N-k)^k$, where $k$ is the number of operators of each term (and $k \sim N$). This path growth provides the scaling of $\tau \sim N^{-\frac{1}{2}}$, which ultimately
gives the advantage.  

However, it is important to understand that Eq.~\eqref{eq:optimal_charging_power_X-model_SYK_2} \textit{is not} a first order expansion for the charging power $\PAvAvX(t)$. Rather, it corresponds to a partial resummation of the \textit{full} series expansion, including among all paths only the dominants ones.
Diagrammatically, the paths considered by this resummation are all the paths where a single Majorana operator is never moved consecutively twice or more. Since the number of Majorana operators in a block of size $k\propto N$ is extensive, the chance of moving a certain operator more than once is overwhelmingly small. This, in turn, explains the almost perfect agreement between Eq.~\eqref{eq:charging_power_X-model_SYK_2} and its full counterpart in Eq.~\eqref{eq:charging_power_X_Z_model}. The same argument clarifies the origin of the excellent agreement between the exact energy stored in the battery, $\overline{E}(t)$, and its approximated counterpart obtained from Eq.~\eqref{eq:charging_power_X-model_SYK_2}, for time scales well-beyond $\tau$ (see the insets in Fig.~\ref{fig:charging_power_SYK_2_full}).

The path-counting heuristic also allows us to rule out extra sources of quantum charging advantage when considering SYK$_q$ Hamiltonians with $q \geq 4$ \textit{but finite}. Indeed, battery Hamiltonians with a size that has no scaling with $N$ will only increase the number of paths \textit{polynomially} with each commutator. Consequently, they provide the same qualitative physics as the $Z$-model and cannot offer a charging advantage \footnote{The impossibility of obtaining a quantum charging advantage for the $Z$-model battery and any finite $q$ can also be obtained as a consequence of the theorem proved in \cite{gyhm2022quantum}.}. When considering the $X$-model, the only quantitative difference will come from how fast the charging power will be achieved; again there is no extra mechanism to realize any charging advantage. Indeed, the very same scaling of charging advantage was obtained numerically in \cite{rossini2021quantum}, for the case of the $X$-model charged via a SYK$_4$ quench. As an aside, we note that a \textit{double-scaling} limit, in which $q$ has some scaling with $N$, is more subtle and could potentially lead to extra charging advantage. We will return to this point in due course. 

\section{SYK in a graph and the Majorana blockade}
From the discussion thus far, it is clear that quantum charging advantage lies in the possibility of simultaneously moving all the operators residing in a \textit{large}, \textit{i.e.} scaling with $N$, block of Majoranas. At the same time, the main obstruction to this mechanism is given by the Pauli exclusion principle for Majoranas, which forbids moving two fermions simultaneously to the same location. For a complete graph, this obstruction is quite immaterial because of the all-to-all nature of the graph. However, when considering graphs with local structure -- for example only nearest-neighbor edges -- this obstruction can dominate the operator dynamics,  destroying the possibility of a charging advantage even for large operators.


\begin{figure}[t!]
    \centering
    \includegraphics[width=0.2\textwidth]{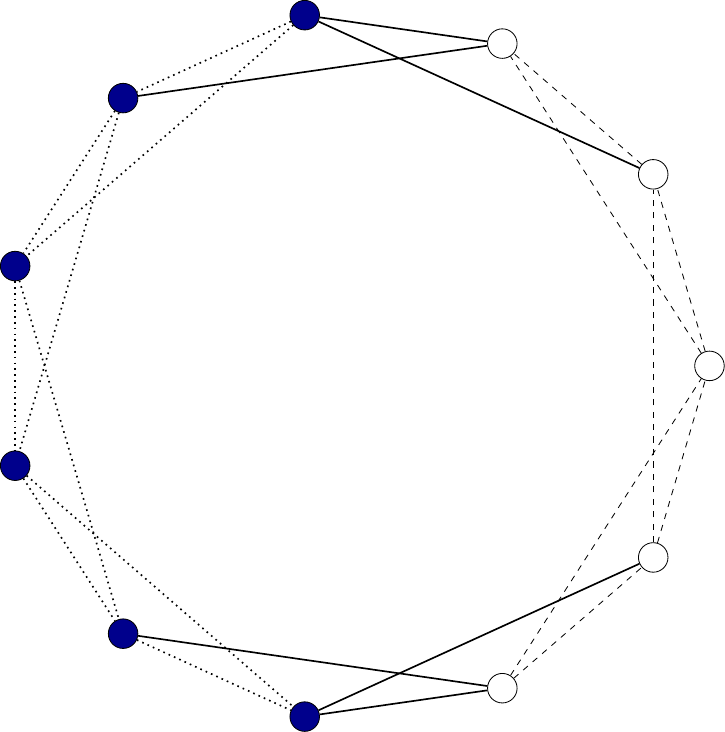}
    \hspace{0.07\textwidth}
    \includegraphics[width=0.2\textwidth]{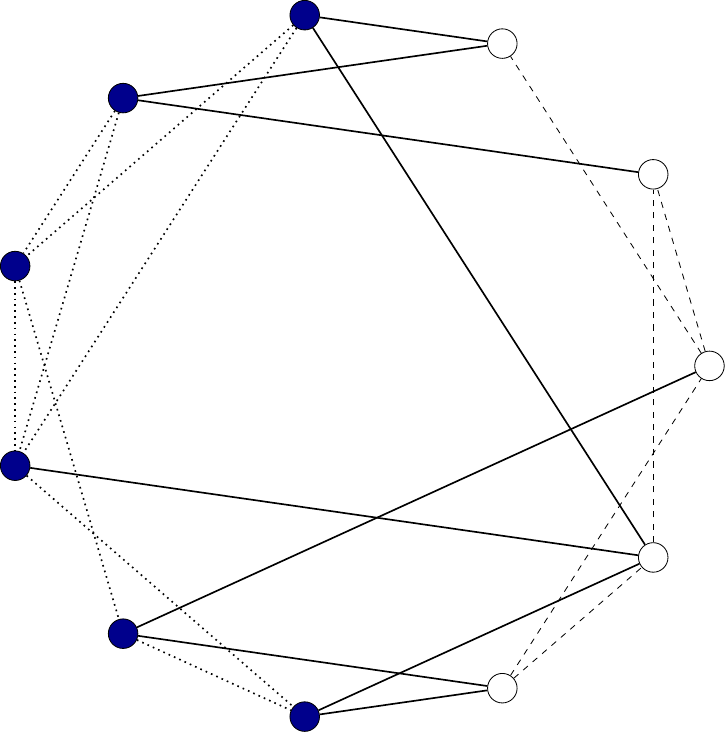}
    \caption{Small-world graphs with $\kappa = 4$,  $N = 11$ and $p = 0$, $p \neq 0 $ in the left and right, respectively. Filled vertices denote occupied Majoranas, solid lines denote possible steps in the random walk and dotted lines denote forbidden steps due to the Majorana blockade.}
    \label{fig:SW_graph}
\end{figure}

Motivated by these considerations, we want to extend the results obtained for the complete case to a broader class of graphs, using the sparse Hamiltonians in Eq.~\eqref{eq:SYK_hamiltonian_graph_def}.  Let us consider a string of $k$ Majoranas in a graph $G$, having an adjacency matrix $A$. We aim to evaluate the average
\begin{equation}
\label{eq:average_k_block_adjacency_matrix}
\overline{\GammaOp_1(t) \ldots \GammaOp_k (t)} = f_A^{(k)}(t) ~ \GammaOp_1 \dots \GammaOp_k \, ,
\end{equation}
by resuming only the dominant contributions in the BCH expansion discussed in the previous section. 

The first step is to identify the relevant paths. 
To this end, we expand $f_A^{(k)}(t)$ at the lowest order to get $ f_A^{(k)}(t) = 1 - \frac{1}{2} g_k t^2 + \cdots$,  where $g_k$ is given by the expression
\begin{equation}
    g_k = \frac{1}{d} \sum_{m=1}^k \sum_{l>k}^N A_{lm},
\end{equation}
and $d$ is the average degree of the graph. Geometrically, the quantity $d \, g_k$ counts the number of edges that go outside of the block $\GammaOp_1 \ldots \GammaOp_k$, \textit{i.e.} the number of first-order, single-Majorana, paths that can delocalize the block $\GammaOp_1 \ldots \GammaOp_k$, represented by solid lines in Fig. \ref{fig:SW_graph}. Hence, we refer to $g_k$ as the \textit{connectivity of the block}. 

The second step consists of summing these lowest-order paths. Let us assume that the quantity $g_k$, for $k \sim N$, has \textit{some scaling} with $N$, \textit{i.e.} that the model does have a quantum advantage that we want to quantify.
If we also assume that the mobility of the block does not change significantly when moving a finite number of Majoranas, \textit{i.e.} that after a finite number of movements, the new connectivity is $g_k' = g_k + O(1)$, we can generalize the previous resummation to get
\begin{equation}
\begin{aligned}
    \label{eq:fk_general_graph}
     f_A^{(k)}(t) &= e^{-\frac{1}{2} g_k t^2} \, \\
     &\Rightarrow \PAvAv_{A} (t) = \frac{N}{2 t} \left( 1 - \sum_{k = 1}^{N/2} e^{- \frac{1}{2} g_{2k - 1} t^2} \right) \, .     
\end{aligned}
\end{equation}

Therefore, we see that when $g_k$, for $k\sim N$, scales with $N$, we obtain a charging advantage.  For a complete graph, $d \approx N$ and $g_k \approx k (1-\frac{k}{N})$, in agreement with the fully connected case. Thus, we have translated a quantum dynamical problem into a path counting problem in the graph, \textit{i.e.} into a fully \textit{graph-theoretical} problem. Eq.~\eqref{eq:fk_general_graph} solves the latter, in terms of a simple property: the connectivity of the block.

Let us now discuss the limits of validity of Eq.~\eqref{eq:fk_general_graph}. 
The derivation is based on two main assumptions: that $g_k$, for $k\sim N$, is scaling with $N$, and that moving a finite number of Majoranas does not change the connectivity of the block up to finite corrections. These two assumptions are necessary to replace $f^k(t)$ with the product of the first-order expansions for the function $f(t)$.

If a certain model does not have a scaling connectivity, the number of paths grows polynomially at each order in the BCH and the optimal charging time $\tau$ has no $N$ scaling. Therefore, no charging advantage is at work in this case. Interestingly, although Eq.~\eqref{eq:fk_general_graph} does not provide a good approximation of the actual charging power, this approximated expression is enough to rule out the charging advantage.

On the other hand, the second assumption is generically violated for graphs that are \textit{hub-dominated}. An extreme example is provided by the star graph (studied in detail in \cite{kim2022operator}), in which all vertices are connected to a single central vertex, with no other connections between nodes. After moving one of the Majoranas to the central node, the graph undergoes a complete blockade and no single Majorana is allowed to move except for the central one. This graph still enjoys a quantum advantage, but our derivation is invalid and an independent path counting must be carried. 

To illustrate the framework, we consider the SYK model defined on small-world graphs, as depicted in Fig. \ref{fig:SW_graph} \cite{kim2022operator}. We build these graphs via the Watts-Strogatz algorithm \cite{watts1998collective}, which uses an integer $\kappa$ and a probability $p$. Starting from a regular circle graph, where each vertex connects to its $\kappa$ nearest neighbors, the algorithm rewires each edge at random with probability $p$, while keeping the graph connected and avoiding repetitions and self-connections.

The geometry of the graph depends critically on $p$. For $p = 0$, we get a chain of Majorana with only local edges. However, when a finite rewiring probability is turned on, we get an order of $\sim p N$ rewirings, yielding a highly non-local graph. In the extreme case $p=1$, the algorithm produces random Erdős-Rényi-like graphs.

\begin{figure}[t!]
    \centering
    \includegraphics[width=0.48\textwidth]{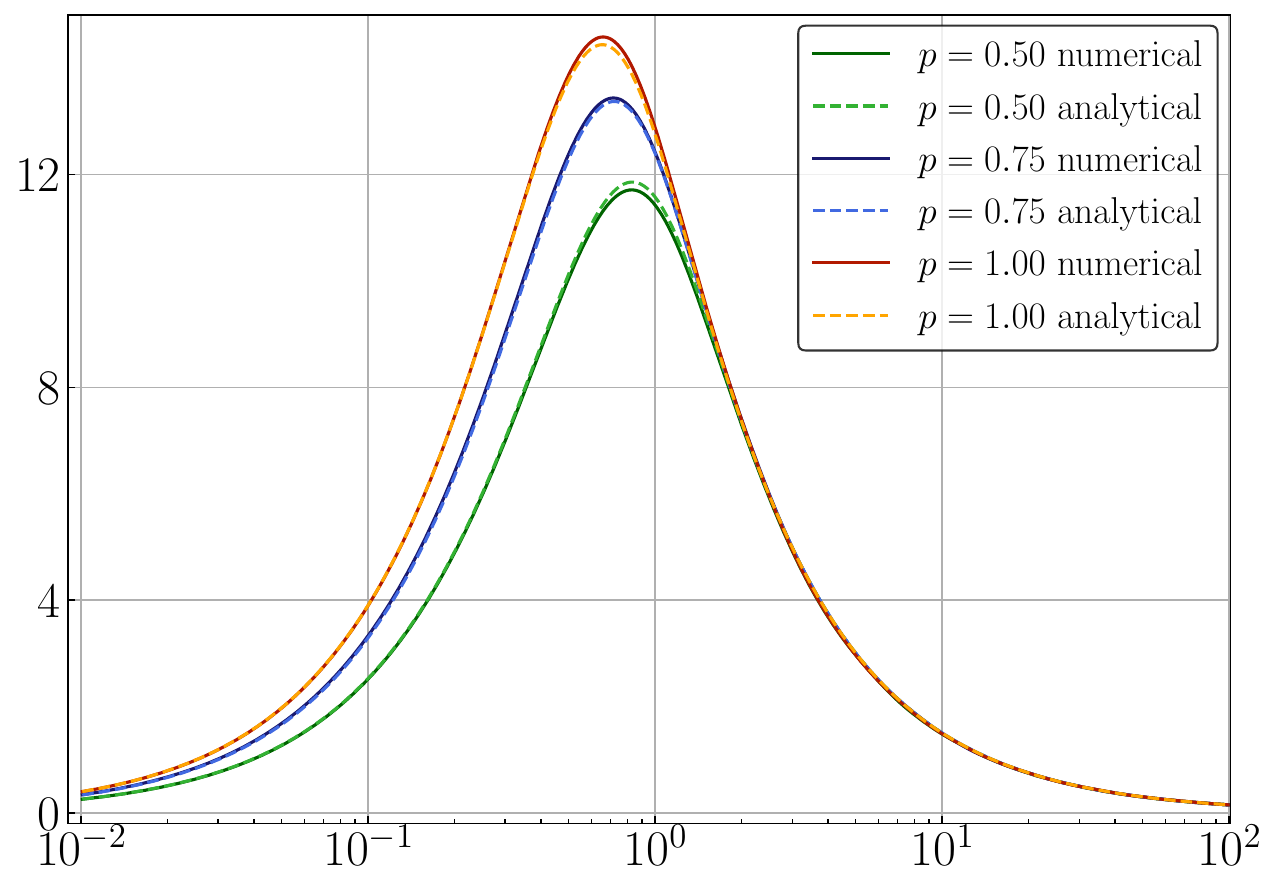}
    \caption{Comparison between the numerical charging power, obtained for $N = 30$ Majorana operators and Watts-Strogatz graphs having $\kappa = 4$ and $p = 0.5, 0.75, 1$, and the analytical expression in Eq.~\eqref{eq:fk_general_graph}.
    }
    \label{fig:charging_SW_graph}
\end{figure}

A simple counting argument shows that, for $p \neq 0$, $\overline{g_{k \sim N}}$ (where the bar denotes an average over graph realizations) scales with $N$. Therefore, from Eq.~\eqref{eq:fk_general_graph}, we conclude that the model \textit{always} enjoys a quantum charging advantage, as long as $p$ is non-vanishing and not inversely scaling with the system size. 
The approximated expression for the charging power, Eq.~\eqref{eq:fk_general_graph}, is tested against its numerical counterparts for Watts-Strogatz graphs having $\kappa = 4$ and various values of $p$. The results, reported in Fig.~\ref{fig:charging_SW_graph}, show excellent agreement for all values of $p$. 


\section{Conclusions}
Quantum batteries present an exciting frontier where foundational concepts in quantum mechanics intersect with practical energy technologies. In this work, we have established an analytical framework for understanding the quantum charging advantage in Sachdev-Ye-Kitaev (SYK) models, systems renowned for their rich connections to quantum chaos, holography, and condensed matter physics. Our analysis identifies the two critical mechanisms for quantum advantage: the use of operators that scale extensively with system size $N$ and the ability of specific graph structures to rapidly delocalize these operators. 

By recasting the charging process as a graph-theoretic problem, we bridge quantum dynamics with graph theory, enabling new insights into how energy transfer in quantum systems depends on structural connectivity. This approach not only demystifies the origins of the quantum charging advantage but also opens a pathway to exploring quantum energy processes in more general contexts, including SYK models on graphs. Our results suggest that graph connectivity and operator size are universal drivers of charging efficiency, with potential implications for a wide range of physical systems.

Beyond quantum batteries, the methods and results presented here resonate with broader themes in high-energy and condensed-matter physics. The random walk interpretation and the consequent partial resummation techniques used to study large operators offer tools for probing quantum thermalization \cite{dalessio2016from}, localization (both at the single particle and the many-body level) \cite{ros2015integrals,abanin2019colloquium, sierant2024man-ybody}, and even the role of operator spreading in holographic dualities. By placing SYK batteries at the nexus of energy storage and foundational physics, we provide a robust platform for future studies exploring the interplay of quantum chaos, operator dynamics, and graph-theoretic structures.

Looking ahead, several intriguing directions emerge from this work. First, extending the graph-theoretic framework to investigate SYK models with more intricate coupling structures or interactions that deviate from Gaussian randomness could yield deeper insights into the universality of the charging advantage. Second, applying this framework to study energy storage and transfer in open quantum systems and non-Hermitian versions of the SYK model \cite{garcia-garcia2022symmetry}, where dissipation and decoherence play a role, could bridge the gap between idealized models and experimental realizations. Third, exploring connections to physical realizations of SYK-like models, such as cold atoms or engineered superconducting circuits \cite{danshita2017creating, pikulin2017black, wei2021optical}, may provide a route to experimentally test the theoretical predictions presented here.

Furthermore, the implications of graph structure and operator dynamics on related phenomena, such as Anderson localization, many-body localization, and entanglement spreading, remain largely unexplored. Investigating these connections could reveal new universality classes or even suggest practical applications in quantum computing and information storage. Finally, adapting the methods developed here to study charging dynamics in systems governed by non-Hermitian or time-dependent Hamiltonians could open the door to novel paradigms for quantum energy systems.

This work represents a step forward in understanding how quantum effects can be harnessed for practical and theoretical advancements, highlighting the power of interdisciplinary approaches to tackling fundamental questions at the intersection of physics and technology.

\begin{acknowledgments}
    We thank Matteo Carrega, Ju-Yeon Gyhm, and William Salazar for their interesting discussions and collaboration on related topics.
    We are particularly grateful to Jan Olle and Jaco Van Zyl for their involvement in the early stages of this project.
    DR thanks the organizers of the ``III Workshop on Quantum Information and Thermodynamics'', where the results of this paper were presented. FD would like to thank the Perimeter/ICTP-SAIFR/IFT-UNESP fellowship program and CAPES for financial support. DR thanks FAPESP, for the ICTP-SAIFR grant 2021/14335-0 and for the Young Investigator grant 2023/11832-9. DR also acknowledges the Simons Foundation for the Targeted Grant to ICTP-SAIFR. JM acknowledges support in part by the “Quantum Technologies for Sustainable Development” grant from the National Institute for Theoretical and Computational Sciences of South Africa (NITHECS). Research at Perimeter Institute is supported in part by the Government of Canada through the Department of Innovation, Science and Economic Development and by the Province of Ontario through the Ministry of Colleges and Universities.
\end{acknowledgments}

\appendix
\section{Jordan-Wigner}
\label{app:JW}

The battery and charging Hamiltonians are formulated in terms of different classes of operators: we have spin operators for $\HOp_0$ and Majorana operators for $\HOp_1$ Hamiltonians. We employ the well-known \textit{Jordan-Wigner transformation} (JW) to connect fermionic with spin operators. It is important to stress that the choice of a specific mapping constitutes part of the charging protocol itself. Therefore, the charging performance of the resulting quantum battery \textit{depends} on this particular choice.

In our case, we use the following JW transformation:
\begin{align}
    \label{eq:JW_def}
    & \GammaOp_{2j - 1} = \frac{1}{\sqrt{2}} \left( \prod_{i = 1}^{j - 1} \PauliOp_i^z \right) \PauliOp_{j}^x \, , \nonumber \\
    & \GammaOp_{2j} = \frac{1}{\sqrt{2}} \left( \prod_{i = 1}^{j - 1} \PauliOp_i^z \right) \PauliOp_{j}^y \, ,
\end{align}
together with \textcolor{darkblue}{its} inverse
\begin{equation}
    \label{eq:JW_inverse_def}
    \PauliOp_i^z = - 2 \im \, \GammaOp_{2 i - 1} \GammaOp_{2 i} \, , \quad \PauliOp_i^x = (\sqrt 2)^{2 i - 1} (- \im)^{i - 1} \prod_{p = 1}^{2 i - 1} \GammaOp_p \, .
\end{equation}

For the analytical derivation of the charging performances, it is more convenient to express the battery Hamiltonians of Eq. (\ref{eq:X_Z_model_def}) to the Majorana fermion: 
\begin{equation}
\begin{aligned}
    & \HOp_0^z = \, -2 \im  \sum_{j=1}^{N/2} \GammaOp_{2j-1} \GammaOp_{2j} , \\
    & \HOp_0^x = \sum_{k=1}^{N/2}  2^{k-\frac{1}{2}} (- \im)^{k+1} \prod_{j=1}^{2k-1} \GammaOp_j \,
\end{aligned}
\end{equation}
From the point of view of the $\HOp_1$, the two batteries are very different: the $Z$-model is given by a sum of size $2$ operators, while the $X$-model is formed by operators of various sizes (up to $N/2$). This marked difference is at the heart of the very different charging performances of the two models.

\end{document}